\documentclass[letter]{article}[11pt]

\usepackage{amssymb, amsmath, setspace, epsfig,graphicx,psfrag,color}

\definecolor{Red}{rgb}{1,0,0}
\definecolor{Blue}{rgb}{0,0,1}
\definecolor{Olive}{rgb}{0.41,0.55,0.13}
\definecolor{Green}{rgb}{0,1,0}
\definecolor{MGreen}{rgb}{0,0.8,0}
\definecolor{DGreen}{rgb}{0,0.55,0}
\definecolor{Yellow}{rgb}{1,1,0}
\definecolor{Cyan}{rgb}{0,1,1}
\definecolor{Magenta}{rgb}{1,0,1}
\definecolor{Orange}{rgb}{1,.5,0}
\definecolor{Violet}{rgb}{.5,0,.5}
\definecolor{Purple}{rgb}{.75,0,.25}
\definecolor{Brown}{rgb}{.75,.5,.25}
\definecolor{Grey}{rgb}{.5,.5,.5}
\definecolor{Pink}{rgb}{1,0,1}
\definecolor{DBrown}{rgb}{.5,.34,.16}

\definecolor{Black}{rgb}{0,0,0}

\long\def\mnew#1{{ #1}}

\newtheorem{claim}{Claim}
\newtheorem{lemma}{Lemma}
\newtheorem{theorem}{Theorem}

\newtheorem{corollary}[claim]{Corollary}
\newenvironment{proof}[1][Proof]{\begin{trivlist}
\item[\hskip \labelsep {\bfseries #1}]}{\end{trivlist}}
\newenvironment{definition}[1][Definition]{\begin{trivlist}
\item[\hskip \labelsep {\bfseries #1}]}{\end{trivlist}}

\newcommand{\qed}{\nobreak \ifvmode \relax \else
      \ifdim\lastskip<1.5em \hskip-\lastskip
      \hskip1.5em plus0em minus0.5em \fi \nobreak
      \vrule height0.75em width0.5em depth0.25em\fi}

\newcommand{\enp} {\hfill \rule{2.2mm}{2.6mm}}

\newcommand{\la}{\lambda}

\newcommand{\ep}{\epsilon}

\newcommand{\mc}{\mathcal}

\newcommand{\mrm}{\mathrm}

\newcommand{\M}{\mathrm{M}}
\newcommand{\PM}{\mathrm{PM}}

\title{Belief-Propagation for Weighted $\mrm{b}$-Matchings on Arbitrary Graphs and its Relation to Linear Programs with Integer Solutions}

\author{Mohsen Bayati\thanks{Microsoft Research;
\{mohsenb~,~borgs~,~jchayes\}@microsoft.com}\and
Christian Borgs$^*$ \and Jennifer Chayes$^*$ \and Riccardo
Zecchina\thanks{Politecnico Di Torino; riccardo.zecchina@polito.it}}

\date{}
\begin{document}

 \maketitle

\begin{abstract}
We consider the general problem of finding
the minimum weight $\mrm{b}$-matching on arbitrary graphs.
We prove that, whenever the linear
programming (LP) relaxation of the problem has no fractional solutions,
then the belief propagation (BP) algorithm converges to the
correct solution.  This result is notable in several
regards:
(1) It is one
of a very small number of proofs showing correctness of
BP without any constraint
on the graph structure.
(2) Instead of showing that BP leads to a PTAS, we give a
finite bound for the number of
iterations after which BP has converged to the exact solution.
(3)  Variants of the proof work for both
synchronous and asynchronous BP; to the best of our knowledge, it is
the first proof of convergence and correctness of an
asynchronous BP algorithm for a combinatorial optimization problem. (4)
It works for both ordinary $\mrm{b}$-matchings and the more difficult case of perfect $\mrm{b}$-matchings.
\mnew{(5) Together with the recent work of Sanghavi, Malioutov and Wilskly \cite{SMW07} they are the first complete proofs
showing that tightness of LP implies correctness of BP.}

\end{abstract}

\section{Introduction}\label{sec:intro}

Motivated by the cavity method in statistical physics, very
fast distributed heuristic algorithms have recently been developed for the solution
of random constraint satisfaction problems \cite{MeZ02},
\cite{BMP03}, \cite{FMV06}, \cite{AcR06}.
Similar heuristic methods have been known for many years
\cite{Gal63} in the context of coding theory.  And a variety of
specific examples of such algorithms have been developed in
artificial intelligence, signal processing, and digital
communications. Well-known examples include the Viterbi
algorithm, the iterative decoding algorithm in turbo codes
and in low-density parity-check codes \cite{RiU01},
Pearl's belief propagation algorithm for Bayesian networks
\cite{Pea88}, the Kalman filter, and certain fast Fourier transform
(FFT) algorithms.  Very recent applications can also be found in
systems biology \cite{Fri04}, \cite{GTR05}, \cite{YaW02},
computer vision \cite{TaF03}, and data clustering \cite{FrD07}.

In some cases, the algorithms generated by the cavity method are
exactly of the form of a classic belief propagation {(max-product or min-sum)}
i.e., a message-passing algorithm for efficiently
computing marginal probabilities or finding the assignment with highest probability of a joint discrete probability
distribution defined on a graph. The belief propagation (BP) algorithm converges to
a correct solution if the associated graph is a tree, and
may be also {a good heuristic}
for some graphs with cycles.
In other cases, the cavity method
may lead to a more involved survey propagation (SP) algorithm \cite{MeZ02},
in which some form of correlation among variables is controlled.

In this paper, we study the problem of finding the minimum weight $\mrm{b}$-matchings in \emph{arbitrary} graphs
via the min-sum version of BP algorithm\footnote{Throughout
this paper, the term BP algorithm refers to min-sum version of the BP algorithm.}.

\paragraph{Our Results.}

Let $G=(V,E)$ be an undirected graph with edge weights $w_{ij}$ for
each edge $\{i,j\}\in E$ and node capacities $b_i$ for each node
$i\in V$. The iterative message-passing algorithm based on synchronous BP for solving the weighted perfect
$\mrm{b}$-matching problem (see our Section \ref{sec:Def-Prob-Stat} for the precise
definition) is the following {\emph{ simple}} procedure:
At each time, every vertex of the graph sends (real {valued}) messages  to each of its neighbors.
The message transmitted at time $t$ from $i$ to $j$ is $w_{ij}$ minus the $b_i^{th}$ minimum of
the messages previously received by
$i$ at time $t-1$ from all of its neighbors except $j$. At the end of each
iteration, every vertex $i$ selects $b_i$ of its adjacent edges that
correspond to the $b_i$ smallest received messages.

We will show the following
result:  For {arbitrary} graphs $G$,
and all sets of weights {$\{w_{ij}\}$}, after $O(n)$
iterations, the set of selected edges converges to the correct solution,
i.e., to the
minimum weight perfect $\mrm{b}$-matching of $G$, provided
that the LP relaxation
of the problem (see Section \ref{sec:Def-Prob-Stat} for definitions)
has no fractional solutions.
Additionally we introduce a new construction, a \emph{generalized computation tree},
which allows us to analyze the more complicated case of
BP with an asynchronous updating scheme, and prove
convergence and correctness of it when each edge of the graph
transmits {at least $\theta(n)$} messages. To the best of our knowledge,
 this technique is new and can be applied in the analysis of
asynchronous BP in other problems as well.
These are extensions of the previous results of \cite{BSS05} and \cite{HuJ07}
which showed convergence and correctness of the above algorithm for
bipartite graphs.\footnote{{Both of these results were assuming that the
minimum weight matching is unique. Note that if there is more than
one solution, then one can
construct a fractional solution to the LP relaxation.}}
Moreover, our proof gives a better understanding of
the often-noted but poorly understood connection
between BP and LP through the dual of the LP
relaxation. We also modify
our BP algorithm and its analysis
to include the problem of finding the \emph{non-perfect}
weighted $\mrm{b}$-matchings. \mnew{Recently and independently from our work a similar result for the scenario of using synchronous BP for non-perfect $1$-matchings was shown by Sanghavi, Malioutov and Wilskey \cite{SMW07}.}

\paragraph{Related Works.}
The weighted $\mrm{b}$-matching problem is an important
problem in combinatorial optimization. It
belongs to a family of integer linear programs which have been
well-studied and can be solved in strongly polynomial time \cite{EdJ70}, \cite{EdK72}.
For extensive surveys see \cite{Ger95} and \cite{Pul95}.
In physics, {the} study of the random $1$-matching problem goes back to the work
of M\`{e}zard and Parisi \cite{MeP86} who made a celebrated conjecture for the expected
optimum weight ($\pi^2/6$) that was proven to be exact a decade later by Aldous \cite{Ald01}.

BP algorithms have been the subject of extensive study in several communities.
The general BP algorithm is known to be correct on graphs with no cycles
\cite{Pea88}. For graphs with a single cycle, convergence and
correctness of BP have also been rigorously analyzed \cite{AHM98},
\cite{Wei00}. For arbitrary graphs, relatively little is known about the
correctness of BP, although some interesting progress has been made in \cite{Wei01},
\cite{WJW04}, \cite{TaJ02}, \cite{YFW00}. Performance of
the BP algorithm usually depends on the length of cycles in graphs;
most analytical results require that the graphs have no short
cycles (i.e., that they are large-girth graphs) \cite{RiU01}, \cite{BaN06},
\cite{GNS05}. For the case of weighted matchings and a
few other problems, there were initially surprising results that
BP works correctly on graphs with many short cycles
(\cite{WeF01}, \cite{RuV01}, \cite{BSS05}, \cite{MoV05}, \cite{MPT06}).

Recent works have also suggested a connection between the BP algorithm
and linear
programming (LP) in particular problems. A relationship
between iterative decoding of channel codes and LP decoding was
studied in \cite{FWK05}, \cite{VoK04}, \cite{VoK05}. Other relationships were noted in the
context of BP algorithms with convex free energies
\cite{WJW05}, \cite{WJW05Upp}, \cite{WYM07}, and in the case of BP algorithms
for resource allocations \cite{MoV07J}. For
weighted $1$-matchings, the connection was studied \cite{BSS06}
in the context of similarities between BP equations and
the primal-dual auction algorithm
of Bertsekas \cite{Ber88}. \mnew{And it was further clarified recently for non-perfect $1$-matchings in \cite{San07} and \cite{SMW07}
where it was shown that BP does not converge to the correct solution if the LP relaxation has fractional
solutions.}
Another recent result studies this connection for the weighted independent set problem
\cite{SSW07}. We will compare our work with some of these results in the
``Technical Contribution'' section below.

Finally, we note that the BP equations for solving the weighted matching problem
which we use in this paper have been previously studied in \cite{BSS06}, \cite{HuJ07}.
These equations are also very similar to equations  {for weighted
matching problems and traveling salesman problems} given in
\cite{MeP86}, \cite{Was06}, \cite{Ald01}, \cite{GNS05}, and to
equations  {for various other problems given} in
\cite{ZdM06}, \cite{PrW06}, \cite{MSV07}.


\paragraph{Technical Contribution.} The main contributions of our results and techniques can be
summarized as follows:
\begin{enumerate}
\item \emph{BP for the weighted matching} was first used in \cite{BSS05}
and its correctness and convergence was shown for bipartite graphs with
unique optimum solution. That proof relied heavily  on the fact
that the minimum weight matching of a bipartite graph is locally optimal
on any cycle since the cycles of a bipartite graph have even length.
The same technique was used in \cite{HuJ07} to extend the result to
$\mrm{b}$-matchings in bipartite graphs. But this technique fails for
graphs containing cycles with odd length. In order to bypass this difficulty
we use a completely different tool, complementary slackness conditions of
the LP relaxation and its dual, which is independent of the graph structure.

\item \emph{Connection between LP and BP} has been suggested and analyzed by various groups
(as we discussed above), but our result \mnew{together with \cite{SMW07}}, to the best of our knowledge, are the first ones which show both
convergence and correctness of the BP algorithm when LP relaxation has no
fractional solutions. One related result, \cite{SSW07}, studies only properties
of the BP fixed points and their relation to the LP, \emph{conditioned} on
the convergence of the BP algorithm. Similarly in another recent work,
\cite{WYM07}, which generalizes methods of \cite{WJW05Upp} and \cite{WJW05},
the connection of the BP algorithm and LP relaxation is studied
{in} the converged case of the BP. {The authors} also study interesting variations
of the BP which have convex free energies.

\item \emph{The asynchronous BP}, which includes the synchronous
version as a special case, has been a more popular version for practical purposes.
But, due to its more complicated structure,
it has not been the subject of much rigorous study. To the best of our
knowledge, our work is the first correctness and convergence proof of asynchronous
BP for a combinatorial optimization problem. Another advantage of our proof is
the construction of a new tool, the generalized computation tree, which can be used
for the analysis of the both convergence and correctness of  asynchronous
message-passing algorithms including BP. Without {the notion of a suitable}
computation tree the existing methods, free energy analysis \cite{YFW00}\cite{WJW04}
or Lipschitz functions \cite{GNS05}\cite{BaN06}, {do not give
correctness and convergence at the same time.}
\end{enumerate}

\paragraph{Organization of the Paper.}

The rest of the paper is organized as follows. In Section
\ref{sec:Def-Prob-Stat}, we provide the setup, define the weighted
$\mrm{b}$-matching problem, and describe the LP relaxation, the dual LP, and
the complementary slackness conditions. In Section
\ref{sec:alg-main-res}, we describe our algorithm for
the minimum weighted perfect $\mrm{b}$-matching problem, and state our
main result.  The analysis of our algorithm is given in Section \ref{sec:analysis}.
The extension of our algorithm and results to the \emph{non-perfect} minimum weighted $\mrm{b}$-matching problem are presented in Section \ref{sec:non-perfect}. Finally, in Section \ref{sec:async-BP}, we state the asynchronous version of the BP algorithm and present its analysis.

\section{Definitions and Problem Statement}\label{sec:Def-Prob-Stat}

Consider an undirected simple graph $G=(V,E)$, with vertices $V =
\{1,\ldots,n\}$, and edges $E$. Let each edge $\{i,j\}$ have
weight $w_{ij}\in \mathbb{R}$. Denote  {the} set of neighbors of each
vertex $i$ in $G$ by $N(i)$. Let $\mrm{b}=(b_1,\ldots,b_n)$ be a
sequence of positive integers such that $b_i\leq deg_G(i)$. A
subgraph $M$ of $G$ is called a \emph{$\mrm{b}$-matching} (\emph{perfect $\mrm{b}$-matching})
 {if the} degree of each vertex $i$ in $M$ is at most $b_i$ (equal to $b_i$).
Denote the set of $\mrm{b}$-matchings (perfect $\mrm{b}$-matchings)
of $G$ by $\M_G(\mrm{b})$ ($\PM_G(\mrm{b})$), and assume that it is non-empty.
 {Clearly} $\PM_G(\mrm{b})\subset \M_G(\mrm{b})$.

The weight of {a (perfect or non-perfect)} $\mrm{b}$-matching
$M$, denoted by $W_M$, is defined by
$W_{M}=\sum_{ij}w_{ij}1_{\{i,j\}\in M}$. In the next two
sections, we will restrict ourselves to the case of \emph{\it
perfect} $\mrm{b}$-matchings. We will extend the analysis to
(possibly non-perfect) $\mrm{b}$-matchings in Section
\ref{sec:non-perfect}. The minimum
weight perfect $\mrm{b}$-Matching ($\mrm{b}$-MWPM), $M^*$, is
defined by \mnew{$M^*=\textrm{argmin}_{M\in \PM_G(\mrm{b})}\ W_{M}$}.
The goal of this paper is to find $M^*$ via a min-sum belief
propagation algorithm. Throughout the paper, we will assume
that $M^*$ is unique.

\paragraph{Linear Programming Relaxation.}
Assigning variables $x_{ij}\in\{0,1\}$ to the edges in
$E$, we can express the   weighted perfect $\mrm{b}$-matching problem
as the problem of finding a vector
$\mrm{x}\in\{0,1\}^{|E|}$ that minimizes the total weight
$\sum_{ij\in E}x_{ij}w_{ij}$, subject to the constraints
$\sum_{{j\in N(i)}}x_{ij}=b_i$ for all $i\in V$.  Relaxing the constraint
that $x_{ij}$ is integer, this leads to the following linear program and its dual:
\begin{equation}
\begin{array}{rcclcrccl}
&&&&&&&&\\
\textrm{min }&&\sum_{\{i,j\}\in E}x_{ij}w_{ij}&&|&\textrm{max }&&\sum_{i=1}^n b_iy_i-\sum_{{\{i,j\}\in E}} \la_{ij}&\\
\textrm{subject to}&&&&|&\textrm{subject to}&&&\\
&&\sum_{j\in N(i)} x_{ij} = b_i&\forall~~ i&|&&&w_{ij}+\la_{ij}\geq y_i + y_j&\forall~~ \{i,j\}\in E\\
&&0\leq x_{ij} \leq 1&\forall~\{i,j\}\in E&|&&&\la_{ij}\geq0&\forall~\{i,j\}\in E\\
&&&&|&&&&\\
&&&&|&&&&\\
&&\textrm{Primal LP}&&|&&&\textrm{Dual LP}&\\
\end{array}
\label{eq:LP-Relax}
\end{equation}

We say the LP relaxation \eqref{eq:LP-Relax} has \emph{no fractional solution} if,
\mnew{every optimal solution $x$ of LP satisfies $x\in\{0,1\}^{|E|}$}.
Note that { absence of fractional solutions implies uniqueness of
integer solutions,
since any convex combination of two integer solutions is a solution to the
LP as well.} We want to show that the
BP algorithm for our problem converges to the
correct solution, provided the LP relaxation (\ref{eq:LP-Relax}) has no fractional solution.

\paragraph{Complementary Slackness Conditions.}
Complementary slackness
for  the LP and its dual state that
the variables $\mrm{x}^{{*}}=(x_{ij}^*)$ and
$\mrm{y}^*=(y_i^*),~\mrm{\la}^{{*}}=(\la_{ij}^*)$ are optimum solutions to the
LP relaxation and its dual (\ref{eq:LP-Relax}),
respectively, if and only if for all edges $\{i,j\}$ of $G$ both $x_{ij}^*(w_{ij}+\la_{ij}^*-y_i^*-y_j^*)=0$.
and $(x_{ij}^*-1)\la_{ij}^*=0$ hold.
See \cite{BoL-book}, \cite{Sch-Book} for more information about LP, dual LP and complementary slackness conditions.

Using the fact that the {LP has no fractional solution}, one can deduce the following modified
complementary slackness conditions: For all $\{i,j\}\in M^*;$ $w_{ij}+\la_{ij}^*=y_i^*+y_j^*$ and for all $\{i,j\}\notin M^*;$ $\la_{ij}^*=0$.

By these conditions and the fact that
 $\la_{ij}^*\geq 0$, we have {that}
$w_{ij}\leq y_i^*+y_j^*$ for all $\{i,j\}\in M^*$, and
$w_{ij}\geq y_i^*+y_j^*$ for all $\{i,j\}\notin M^*$. {However, as the counterexample given in Appendix \ref{sec:counter-example}
shows, it is in general
not true that these inequalities are strict even when
the LP has no fractional solution.
Let $S$ be the set of those edges in $G$ for
which $|w_{ij}-y_i^*-y_j^*|>0$. We will assume the minimum gap is
$\epsilon$. i.e. $\epsilon = \min_{\{i,j\}\in S}~|w_{ij} - y_i^*-y_j^*|>0$. Throughout this paper we assume that there exist an edge in $G$ for which
the strict inequality $|w_{ij}-y_i^*-y_j^*|>0$
holds and therefore $\epsilon>0$ is well defined. The other cases,
where for each $\{i,j\}\in E$ the equality $w_{ij}=y_i^*+y_j^*$ holds,
happens only for special cases and are discussed in Section \ref{sec:epsilon=0}.
Let also $L = \max_{1\leq i\leq n}~|y_i^*|$ .


\section{Algorithm and Main Result}\label{sec:alg-main-res}

The following algorithm is  {a} synchronous implementation of BP for
finding the minimum weight perfect $\mrm{b}$-matching ($\mrm{b}$-MWPM). The main intuition behind this
algorithm (and, indeed, all BP algorithms) is that each vertex of the graph
assumes the graph has no cycles, and makes the best (greedy) decision
based on  {this} assumption. This is shown in more detail in Section
\ref{subsec:BP-solv-comp-tree}.


Before applying the BP algorithm, we remove all \emph{trivial} vertices from the graph.
A vertex $i$ is called trivial if $deg_G(i)=b_i$. This is because all of the edges adjacent to $i$ should be in every perfect $\mrm{b}$-matching. Therefore the graph can be simplified by removal of all trivial vertices and their adjacent edges.
\vspace{.1in}
\hrule
\noindent{\bf Algorithm Sync-BP.}
\vspace{2mm}
\hrule
\begin{itemize}
\item[(1)] At times $t=0,1,\ldots$, each vertex sends  {real-valued} messages
to each of its neighbors.  {The} message of $i$ to
$j$ at time $t$ is denoted by $m_{i\to j}(t)$.

\item[(2)] Messages are initialized\footnote{We show in Section \ref{sec:indep-initial} that the messages can be initialized by any arbitrary values.} by $m_{i\to j}(0)
= w_{ij}$ for all $\{i,j\}\in E$.

\item[(3)] For $t \geq 1$, messages in iteration $t$ are obtained from
messages  {in} iteration $t-1$ recursively as follows:
\begin{eqnarray}
\forall~\{i,j\}\in E:~~~~m_{i\to j}(t) & = & w_{ij} -
b_i^{th}\textrm{-min}
_{\ell\in N(i)\backslash\{j\}}\bigg[
m_{\ell\to i}(t-1)\bigg] \label{l:recursesim}
\end{eqnarray}
where $k^{th}$-min$[A]$  {denotes} the $k^{th}$ minimum\footnote{Note that the $b_i^{th}\textrm{-min}_{\ell\in N(i)\backslash\{j\}}$ is well defined since we assumed that all trivial vertices are removed and thus there are at least $b_i+1$ elements in the set $N(i)$ for each $i$.} of set A.

\item[(4)] The estimated $\mrm{b}$-MWPM at the end of iteration $t$ is $M(t)=\cup_{i=1}^nE_i(t)$ where $E_i(t)=\big\{\{i,j_1\},\ldots,\{i,j_{b_i}\}\big\}$  is such
that $N(i)=\{j_1,j_2,\ldots,j_{deg_G(i)}\}$ and $m_{{j_1}\to i}(t)
\leq m_{{j_2}\to i}(t)\cdots\leq m_{{j_{deg_G(i)}}\to i}(t)$.
i.e.,  among all $i$'s neighbors, choose edges to the $b_i$
neighbors that transfer
the smallest incoming messages to $i$.

\item[(5)] Repeat (3)-(4)  {until} $M(t)$
    converges\footnote{{ The subgraph $M(t)$ is not
    necessarily a perfect $\mrm{b}$-matching of $G$ but we
    will show that after $O(n)$ iterations it will be the
    minimum weight perfect $\mrm{b}$-matching.}}.
\end{itemize}
\vspace{2mm}
\hrule
In Corollary \ref{cor:bp-solves-tree}, we will show the main intuition behind
the equation \eqref{l:recursesim} and how it is derived. But we note that one can also
use the graphical model representations of \cite{BSS05}, \cite{HuJ07},
\cite{San07} to obtain the standard BP equations for this problem,
which, after some algebraic calculations, yield the recursive
equation \eqref{l:recursesim}.

{The main result of the paper is rather surprising:  it
says that the above algorithm, which is designed for graphs with no
cycle (i.e., for trees), works correctly for a much larger family of graphs
including those with many short cycles.}

\begin{theorem}\label{thm:main}
Assume that {the LP relaxation (\ref{eq:LP-Relax})
has no fractional solution}.  {Then} the algorithm Sync-BP converges to $M^*$ after
at most $\lceil \frac{2nL}{\epsilon}\rceil$ iterations.
\end{theorem}
%

{If the LP relaxation \eqref{eq:LP-Relax} has a fractional solution
whose cost is strictly less than $W_{M^*}$, then \cite{San07}, \cite{SMW07} have shown
for the case of 1-matching that BP does not converge to $M^*$.
It is straightforward to generalize this to
perfect $\mrm{b}$-matching as well.
But for the case {in which} the LP relaxation has
a fractional solution whose cost is equal to $W_{M^*}$, BP fails
in general. This is because the $b_i^{th}$ minimum in
equation \eqref{l:recursesim} is not unique, and one needs
an oracle to make the right decision. If such an oracle {exists},
then BP converges to $M^*$.}

In Section \ref{sec:non-perfect}, we will give and analyze
{the analogous min-sum algorithms for} 
finding the (possibly non-perfect) minimum weight $\mrm{b}$-matching and in Section \ref{sec:async-BP}, we will give and analyze the asynchronous version of Sync-BP.

\section{Analysis of the Synchronous BP}\label{sec:analysis}

This section {contains the analysis of the synchronous BP algorithm
for perfect b-matchings.} First, in Section
\ref{subsec:BP-solv-comp-tree} we show one derivation of
the equations for Sync-BP and its representation in term of the so-called computation
tree. Next, in Section \ref{subsec:tech-lemma} we introduce a technical
lemma {which connects the} complementary slackness conditions of Section
\ref{sec:Def-Prob-Stat} with alternating paths in the graph $G$. This
lemma is used in Section \ref{subsec:comp-tree=graph} to prove that,
{when the LP relaxation has {no fractional} solutions, then
solutions on the computation tree are the same as the solutions
on the original graph $G$.}

\subsection{{Computation Tree and Derivation of Sync-BP}}
\label{subsec:BP-solv-comp-tree} The main idea behind the
algorithm Sync-BP is that it assumes the graph $G$ has no
cycle. {In other words,} it finds the $\mrm{b}$-MWPM of a graph $G'$
{that has the same local structure as $G$ but no cycles.}
In this section we
rigorously define such graph $G'$ (computation tree) and show
its connection with the Sync-BP algorithm.

\paragraph{Computation Tree.}
For any $i\in V$, let $T_{i}^t$ be the $t$-level computation
tree corresponding to $i$, defined as follows: $T_{i}^{t}$ is a
weighted tree of height $t+1$,
{rooted at $i$}. All tree-nodes have labels
from the set $\{1,\ldots,n\}$ according to the following
recursive rules:

(a) {The root} has label $i$.

(b) The {set of labels of the} $deg_G(i) $ children of the root is
equal to
$N(i)$.

(c) {If $s$ is a non-leaf node whose parent has label $r$,
then the set of labels of its children is $N(s)\backslash\{r\}$.}
%
%
%
%

\noindent{\bf Note 1.}
$T_{i}^{t}$ is often called the {\em
{unwrapped} tree} at node $i$. The computation tree is
constructed by replicating the local connectivity of the original
graph. The messages received by node
$ i$ in the belief propagation algorithm after $t$ iterations in graph
$G$ are equivalent to those that would {have been} received by the root $ i$
in the computation tree, if the messages {were} passed up along the
tree from the leaves to the root. Computation trees have been used in
most of the previous {analyses} of BP algorithms; see e.g.
\cite{Gal63, BSS05, Wei00, Wei01, WeF01, FrK00}.

A subtree $\mc{M}$ of edges in the computation tree $T_{
i}^{t}$ is called a perfect \emph{tree-$\mrm{b}$-matching} if
for each \emph{non-leaf} vertex with label $i$ we have
$deg_{\mc{M}}( i)=b_i$. Now denote the minimum weight perfect
tree-$\mrm{b}$-matching ($\mrm{b}$-TMWPM) of the computation
tree $T_{i}^{t}$ by $\mc{N}^*(T_{ i}^{t})$. We will show that
Sync-BP {can be seen as a dynamic programming procedure}
{that}
finds the minimum weight perfect tree-$\mrm{b}$-matching over
the computation tree. Figure \ref{fig:one} shows a graph $G$
and one of its corresponding computation tree.

\begin{figure}
\centering
      \includegraphics[scale=0.5]{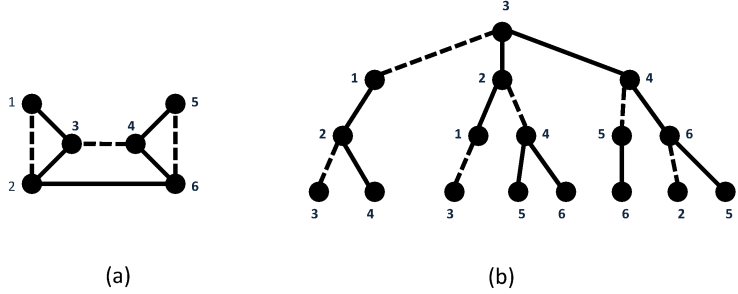}
      \caption{ Part (a) shows a graph $G$ where dashed and gray edges represent a $1$-matching. Part (b) shows the computation tree $T_3^2$ corresponding to $G$ where the set of dashed and gray edges form a $1$-TMWPM.}
      \label{fig:one}
\end{figure}

\paragraph{{Sync-BP Equations.}}
Consider the computation tree $T_{ i}^{t}$. Let us assume that
$deg_G( i)=k$, and {that} $ {i_1},\ldots, {i_k}$ are neighbors of $ i$
in $G$ which are children of the root $ i$ as well. Let us denote
the subtree of $T_{ i}^{t}$ that consists of the root edge
$( i, {i_j})$ and all descendants of $ {i_j}$ by $T_{ {i_j}\to
 i}^{t}$.  {Given this, we define the following weights and
 weight differences:}
\begin{eqnarray*}
W_{ {i_j}\to  i}^+(t)&=&\textrm{Weight of $\mrm{b}$-TMWPM
in $T_{ {i_j}\to  i}^{t}$ that contains the root edge $( i, {i_j})$}.\\
W_{ {i_j}\to  i}^-(t)&=&\textrm{Weight of $\mrm{b}$-TMWPM
in $T_{ {i_j}\to  i}^{t}$ that \emph{does not} contain the root edge $( i, {i_j})$}.\\
n_{ {i_j}\to  i}(t)&=&W_{ {i_j}\to  i}^+(t)-W_{ {i_j}\to
 i}^-(t).
\end{eqnarray*}
{Clearly, for} any edge $\{i,j\}$ of graph $G$ the real number
$n_{{j\to i}}(t)$ is well-defined; the next lemma shows its relation with the messages passed
in  Sync-BP.
\begin{lemma}\label{lem:m=n} For all $1\leq i,j\leq n$ such that $\{i,j\}$
is an edge of $G$ and all $t=0,1,\ldots$, the following is true: $
n_{ j\to  i}(t)=m_{ j\to  i}(t)$.
\end{lemma}
\begin{proof}
{We proceed} by induction on $t$. For $t=0$ by definition the
computation tree $T_{ i}^{0}$ has height $1$. Therefore each branch
$T_{ {i_j}\to  i}^{0}$ consists of a single root edge
$( i, {i_j})$. Thus $W_{ {i_j}\to  i}^+(0)= w_{ii_j}$ and
$W_{ {i_j}\to  i}^-(0)= 0$ which gives: $n_{ {i_j}\to
 i}(0)=w_{ii_j}$, and by definition this is equal to $m_{ {i_j}\to
 i}(0)$. Now for the general case consider the computation tree
$T_{ i}^{t}$ and fix a branch $T_{ {i_j}\to  i}^{t}$. Denote the
children of $ {i_j}$ in this branch by $ {j_1},\ldots, {j_\ell}$
with $\ell = deg_G( {i_j})-1$ (by rule (c) from the construction of
the computation tree described above). For simplicity of notation let
$a=b_{i_j}$. Without loss of generality assume that the children
$ {j_1},\ldots, {j_\ell}$ are ordered so that
\[
W_{ {j_1}\to  {i_j}}^+(t-1)\leq W_{ {j_2}\to  {i_j}}^+(t-1) \leq
\cdots \leq W_{ {j_\ell}\to  {i_j}}^+(t-1).
\]
Now it is not hard to see that
\begin{eqnarray*}
W_{ {i_j}\to  i}^+(t)&=&w_{ii_j}+\sum_{r=1}^{a-1}W_{ {j_r}
\to  {i_j}}^+(t-1)+\sum_{r=a}^{\ell}W_{ {j_r}\to  {i_j}}^-(t-1)\\
W_{ {i_j}\to  i}^-(t)&=&\sum_{r=1}^{a}W_{ {j_r}\to
 {i_j}}^+(t-1)+\sum_{r=a+1}^{\ell}W_{ {j_r}\to  {i_j}}^-(t-1),
\end{eqnarray*}
{so that}
\begin{eqnarray*}
n_{ {i_j}\to  i}(t)&=&W_{ {i_j}\to  i}^+(t)-W_{ {i_j}\to  i}^-(t)\\
&=&w_{ii_j}+W_{ {j_a}\to  {i_j}}^-(t-1)-W_{ {j_{a+1}}\to  {i_j}}^-(t-1)\\
&=&w_{ii_j}-n_{ {j_a}\to  {i_j}}(t-1)\\
&=&w_{ii_j}-a^{th}\textrm{-min}_{ r\in
N( {i_j})\backslash\{ i\}}\bigg( n_{ {j_r}\to  {i_j}}(t-1)\bigg).
\end{eqnarray*}
Therefore we have shown that variables $n_{ j\to  i}(t)$ satisfy
the same recursive relation as variables $m_{ j\to  i}(t)$,
equation (\ref{l:recursesim}), and satisfy the same initial
conditions. Thus they are equal.\enp
\end{proof}

It follows immediately from the above lemma that the set of edges $E_i(t)$ which is selected in iteration
$t$ of the algorithm Sync-BP consists of exactly the same edges
{which} are adjacent to root $i$ in $\mc{M}^*(T_i^t)$.
{This is formalized in the following corollary.}
\begin{corollary}\label{cor:bp-solves-tree}
The algorithm Sync-BP solves the $\mrm{b}$-TMWPM problem on the computation
tree. In particular, for each vertex $ i$ of $G$, the set of
$E_i(t)$ {which} was chosen at the end of iteration $t$ by
Sync-BP is exactly the set of $b_i$ edges {which} are attached to the
root in $\mrm{b}$-TMWPM of $T_{ i}^{t}$.
\end{corollary}
Corollary \ref{cor:bp-solves-tree} characterizes the estimated $\mrm{b}$-MWPM,
 $M(t)$, and will be used in the proof of the main result in
 Subsection \ref{subsec:comptree=graph}. In the next subsection
 we present a lemma that is crucial for the proofs of Subsection \ref{subsec:comptree=graph}.

\subsection{Main Technical Lemma}\label{subsec:tech-lemma}
In this section we {state} our main technical lemma
{which} connects the complementary slackness conditions from
Section \ref{sec:Def-Prob-Stat} to paths on the graph $G$ and
on the computation tree. This lemma is a key step in our proof.
Its proof is quite delicate, and provides the connection between the absence of fractional
solutions and the correctness of BP.

\begin{definition}\label{def:alt-path}
A path $P=( {i_1}, {i_2},\ldots, {i_k})$ in $G$ is called
\emph{alternating path} if:
\begin{itemize}
\item[(a)] There exist a partition {of} edges of
$P$ into two sets $A,B$ such that either $
(A\subset M^*~,~B\cap M^*
=\emptyset)$ or $(A\cap M^*=\emptyset~,~B\subset M^*)$.
Moreover $A$ ($B$) consists of all odd (even) edges; i.e.
$A=\{( {i_1}, {i_2}), ( {i_3}, {i_4}),\ldots\}$
($B=\{( {i_2}, {i_3}), ( {i_4}, {i_5}),\ldots\}$).

\item[(b)] The path $P$ might intersect itself or even
repeat its own edges but no edge is repeated immediately.
That is, for any $1\leq r\leq k-2:~~~~ i_r\neq  i_{r+1}$ and $ i_r\neq  i_{r+2}$.
\end{itemize}
$P$ is called an \emph{alternating cycle} if $ {i_1}= {i_k}$.
\end{definition}
\begin{lemma}\label{lem:path-2n}
{Assume that the LP relaxation (\ref{eq:LP-Relax}) has no fractional}
{solution.}
Then for any alternating path $P$ of length at least $2n$, there
exists an edge $\{i,j\}\in P$ such that
{the} inequality $|w_{ij} -
y_i^*-y_j^*|>0$ holds. That is, $P\cap S\neq\emptyset$.
\end{lemma}

\begin{proof} We will consider two cases:

\emph{Case I) Existence of an even simple cycle in $P$.}\\
Consider the subgraph of $G$ that is generated by edges and vertices
of $P$. If this subgraph contains an alternating cycle $C$ that does
not intersect itself (simple cycle) and has even length, then we will
show that $C\cap S\neq\emptyset$. Let
$C=( {j_1},\ldots, {j_{2\ell}}, {j_1})$. Without loss of
generality assume that odd edges belong to $M^*$ and even edges do
not. That is, for all $1\leq r\leq \ell:$
\[\{ {i_{2r-1}},  {i_{2r}}\}\in M^*~~~,~~~\{ {i_{2r}},  {i_{2r+1}}\}\notin M^*\]
{where} $ {j_{2\ell+1}}= {j_{1}}$. To prove $C\cap S\neq\emptyset$,
assume the contrary; that is, assume for all edges $\{i,j\}$ of
$C:~  w_{ij} = y_i^*+y_j^*$. The weight of $M^*$-edges
of $C$ is equal to weight of their complement in $C$, {due to the fact
that}
\[
\sum_{r=1}^{\ell}w_{j_{2r}j_{2r+1}}
=\sum_{s=1}^{2\ell}y_r^*=\sum_{r=1}^{\ell}w_{j_{2r-1}j_{2r}}.
\]
Now one can obtain a perfect $\mrm{b}$-matching $M'$ in $G$ {which} is
different from $M^*$ and has the same weight as $M^*$. This can be
done by defining $M'=M^*$ outside cycle $C$, and $M'=C\backslash M^*$
on cycle $C$. {However,} this contradicts the uniqueness assumption for
$\mrm{b}$-MWPM in $G$ {which holds due to the fact that the LP relaxation has no fractional solution}. {Hence} we are done.

\emph{Case II) There is no even simple cycle in $P$.}\\
Let $P=\{ {i_1}, {i_2},\ldots, {i_k}\}$. Since $P$ has length at
least $2n$, {it must} repeat a vertex. We also add a
natural direction to each edge $\{ {i_j}, {i_{j+1}}\}$ that is
from $ {i_j}$ to $ {i_{j+1}}$. Consider the first vertex that is
revisited by starting from $ {i_1}$ and walking along $P$. That is,
consider the smallest numbers $r,s$ such that $1\leq r<s\leq n+1$
and $ {i_r}= {i_s}$. Now we break $P$ into three connected pieces
as follows:
\begin{itemize}
\item[(i)] Simple path $P_0=( {i_1}, {i_2},\ldots, {i_r})$
(this part will be ignored).
\item[(ii)] Simple cycle $C_1=( {i_r}, {i_{r+1}},\ldots, {i_{s}})$.
\item[(iii)] Path $P_1=( {i_{s+1}}, {i_{s+1}},\ldots, {i_k})$.
\end{itemize}
From now on we are going to assume that path $P_0$ does not even
exist. Basically we will show that there is one edge from $S$ {which}
is in $C_1\cup P_1$. Since we assumed that $P$ has no even simple
cycle, it follows that $C_1$ has odd length ($s-r$ is odd). Since the
length of $P$ is at least $2n$, {it follows that}
$P$ has to intersect itself at
least twice and there must be another vertex that is revisited after
$ {i_r}$. Consider the smallest numbers $r',s'$ such that $r\leq
r'<s'\leq k$ and $ {i_{r'}}= {i_{s'}}$. Denote this new simple
cycle by $C_2$; i.e., $C_2=( {i_{r'}}, {i_{r'+1}},\ldots, {i_{s'}})$.
Again since $C_2$ is an alternating path, it has to have odd length
($s'-r'$ is odd).

Now we claim that $s\leq r'$. Again assume the contrary, that $r<
r'< s$. We obtain a contradiction by finding an even simple cycle in
$P$. Break path $C_1$ in two simple paths
$Q_1=( {i_r}, {i_{r+1}},\ldots, {i_{r'}})$ {and}
$Q_2=( {i_{r'}}, {i_{r'+1}},\ldots, {i_s})$, and define the simple
path $Q_3=( {i_{s}}, {i_{s+1}},\ldots, {i_{s'}})$. Now consider
the simple cycle $C_3=Q_1\cup Q_3$. The length of $C_3$ {is equal} to
$r'-r+s'-s$, which has the same parity as $s-r+s'-r'$, which is even.
Therefore $C_3$ is an even cycle. Moreover, {the fact that
the parities of $r'$ and $s'$
are} different guarantees the alternation of adjacent edges
$\{ {i_{r'-1}}, {i_{r'}}\}$ and $\{ {i_{s'-1}}, {i_{s'}}\}$ in
cycle $C_3$. Similarly the difference in parity {between} $r$ and $s$
implies alternation of adjacent edges $\{ {i_{r}}, {i_{r+1}}\}$ and
$\{ {i_{s}}, {i_{s+1}}\}$ in cycle $C_3$. Thus $C_3$ is an even
length alternating simple cycle, which is a contradiction. So
{the} claim
$s\leq r'$ is proved.

Now we are left with a final possibility which uses the integrality of
{the} LP optimum solution.
Consider the following three pieces of path $P$:
\begin{itemize}
\item[(i)] Simple odd cycle $C_1$.
\item[(ii)] Simple path $P_2=( {i_{s+1}}, {i_{s+1}},\ldots, {i_{r'}})$
(could be only a point).
\item[(iii)] Simple odd cycle $C_2$.
\end{itemize}
If $(C_1\cup P_2\cup C_2)\cap S=\emptyset$, this means {that} for all edges
$\{i,j\}\in C_1\cup P_2\cup C_2$, {the} equality $w_{ij}=y_i^*+y_j^*$
holds. We will reach a contradiction by showing the existence of an
optimum fractional solution for LP relaxation (\ref{eq:LP-Relax}).
This is done by defining $\mrm{x'}$ as follows:
\[
\forall~\{i,j\}\in E:~~~~~~~x_{ij}'= \left\{
\begin{array}{ll}
x_{ij}^*&\textrm{ if } \{i,j\}\notin C_1\cup P_2\cup C_2\\
1-x_{ij}^*&\textrm{ if } \{i,j\}\in P_2\\
0.5&\textrm{ if } \{i,j\}\in C_1\cap C_2   .
\end{array}
\right.
\]
First we need to show that $\mrm{x'}$ is a feasible solutions for
{the} LP.
{For this,} all we need to show is that $\mrm{x'}$ satisfies the same local
constraints as $\mrm{x^*}$ on vertices of $C_1\cup P_2\cup C_2$.
Since all $C_1\cup P_2\cup C_2$ is a connected alternating path, then
for all vertices $ {i_\ell}\in C_1\cup P_2\cup C_2$ ($\ell\notin
\{r,s,r',s'\}$) it is clear that
$x_{(\ell-1)\ell}'+x_{\ell(\ell+1)}'=x_{(\ell-1)\ell}^*+x_{\ell(\ell+1)}^*=1$.
For $\ell = r$,  using the fact that length of $C_1$ is odd and path
$C_1\cup P_2$ is an alternating sub-path of $P$, either
$x_{r(r+1)}^*=x_{(s-1)s}^*=1,~x_{s(s+1)}^*=0$ or
$x_{r(r+1)}^*=x_{(s-1)s}^*=0,~x_{s(s+1)}^*=1$, which leads to
$x_{r(r+1)}'=x_{(s-1)s}'=0.5,~x_{s(s+1)}'=1$ or
$x_{r(r+1)}'=x_{(s-1)s}'=0.5,~x_{s(s+1)}'=0$, respectively. In both
cases, $\mrm{x'}$ satisfies same local constraint as $\mrm{x^*}$ at
$ {i_r}$. {A similar argument holds at} $ {i_{r'}}$.

{Next} we show that $\mrm{x'}$ has the same cost as $\mrm{x^*}$. This
is done by applying the equality $w_{ij}=y_i^*+y_j^*$ to all edges of
$C_1\cup P_2\cup C_2$ as follows:
\begin{eqnarray*}
\sum_{\{i,j\}\in C_1\cup P_2\cup C_2} w_{ij}x_{ij}^*
&=&\sum_{ i\in C_1\cup P_2\cup C_2}y_i^* + x_{i_{r}i_{r+1}}^*y_{i_r}^*
+ x_{i_{r'}i_{r'+1}}^*y_{i_{r'}}^*\\
&=&\sum_{ i\in C_1\cup C_2} y_i^*+\sum_{\{i,j\}\in P_2} w_{ij}x_{ij}'\\
&=&\sum_{\{i,j\}\in C_1\cup C_2\cup P_2} w_{ij}x_{ij}'.
\end{eqnarray*}
This {completes the} proof of Lemma \ref{lem:path-2n}. \enp
\end{proof}

\subsection{Proof of Theorem \ref{thm:main}}\label{subsec:comptree=graph}
\label{subsec:comp-tree=graph}
We will prove Theorem \ref{thm:main}, namely that {if}
 {the LP relaxation (\ref{eq:LP-Relax})  has no fractional solution and hence $M^*$ is unique}, then  Sync-BP
converges to the correct $\mrm{b}$-MWPM. We will do this by
showing that if the depth of computation tree is large enough, then for
any vertex $i$, its neighbors in $M^*$ ($\mrm{b}$-MWPM of $G$) are
exactly those children that are selected in $\mc{N}^*(T_{ i}^{t})$
($\mrm{b}$-TMWPM of $T_{ i}^{t}$). Here is the main lemma that
summarizes the above claim:
\begin{lemma}\label{lem:tree=graph}
If the LP relaxation (\ref{eq:LP-Relax}) has no fractional solution, then
for any
vertex~$i$ of $G$ and for any $t> \frac{2nL}{\epsilon}$, the set of
edges that are adjacent to root $i$ in $\mc{N}^*(T_{ i}^t)$ are
exactly those edges that are connected to $ i$ in $M^*$.
\end{lemma}
{The} proof of Lemma \ref{lem:tree=graph} is the main technical part of
this paper. {Before entering into the details of the proof} here is a high
level overview of the underlying argument.  Consider the computation tree
($T_{ i}^t$) rooted at vertex $ i$ and look at
$\mc{N}^*(T_{ i}^t)$. We will assume that the claim of the lemma
does not hold. That is, we assume that at the root,
$\mc{N}^*(T_{ i}^t)$ does not choose the same edges as $M^*$-edges
adjacent to $ i$. Then we use the property of perfect tree-$\mrm{b}$-matchings,
{namely}
that each non-leaf vertex $ j$ is connected to exactly $b_j$ of its
neighbors, to construct a new perfect tree-$\mrm{b}$-matching on the computation
tree. This new perfect tree-$\mrm{b}$-matching is going to have less total
weight if the depth of the computation tree is large enough. This last
step uses an {alternating path argument which is a highly non-trivial generalization
of the technique of \cite{BSS05} for the case of perfect $1$-matching in
bipartite graphs. For this part we will use the solutions to the dual LP
(\ref{eq:LP-Relax}).

\begin{proof}[Proof of Lemma \ref{lem:tree=graph}]
Let us denote the lifting of a perfect $\mrm{b}$-matching $M^*$ to a perfect tree-$\mrm{b}$-matching
on $T_{ i}^t$ by $\mc{M}^*$. That is, $\mc{M}^*$
consists of all edge of the computation tree with endpoint labels
$ i, j$ such that $\{i,j\}\in M^*$ as an edge in $G$. The goal
is to show that $\mc{N}^*(T_{ i}^t)$ and $\mc{M}^*$ have the same
set of edges at the root of the computation tree. To lighten the notation,
we denote the  $\mrm{b}$-TMWPM of  $T_{ i}^t$  {by} $\mc{N}^*$.

Assume the contrary, that there exist children $ {i_{-1}}, {i_1}$
of root $ i$ such that $\{ i, {i_1}\}\in
\mc{M}^*\backslash\mc{N}^*$ and $\{ i, {i_{-1}}\}\in
\mc{N}^*\backslash\mc{M}^*$. Since both $\mc{M}^*,~\mc{N}^*$ are
perfect tree-$\mrm{b}$-matchings, they have $b_{i_1}$ edges connected to
$i_1$. Therefore there exist a child $ {i_2}$ of $ {i_1}$ such
that $\{ {i_1}, {i_2}\}\in \mc{N}^*\backslash\mc{M}^*$. Similarly
there is a child $ {i_{-2}}$ of $ {i_{-1}}$ such that
$\{ {i_{-1}}, {i_{-2}}\}\in \mc{M}^*\backslash\mc{N}^*$. Therefore
we can construct a set of alternating paths $P_{\ell}, ~\ell \geq 0$,
in the computation tree, that contain edges from $\mc{M}^*$ and
$\mc{N}^*$ alternatively defined as follows. Let $ {i_0} =
\mbox{root}~ i$ and $P_0 = ( {i_0})$ be a single vertex
path. Let $P_1 = ( {i_{-1}},  {i_0},  {i_1})$, $P_2 =
( {i_{-2}},  {i_{-1}},  {i_0},  {i_1},  {i_2})$ and similarly
for $r \geq 1$, define $P_{2r+1}$ and $P_{2r+2}$ recursively as
follows:
\[
P_{2r+1} = ( {i_{-(2r+1)}}, P_{2r},  {i_{2r+1}})
~~~~,~~~~
P_{2r+2} = ( {i_{-(2r+2)}}, P_{2r+1},  {i_{2r+2}})
\]
where $ {i_{-(2r+1)}},~ {i_{2r+1}} $ are nodes at level $2r+1$
such that $\{ {i_{2r}}, {i_{2r+1}}\}\in
\mc{M}^*\backslash\mc{N}^*$ and $\{ {i_{-2r}}, {i_{-(2r+1)}}\}
\in \mc{N}^*\backslash\mc{M}^*$. Similarly
$ {i_{-(2r+2)}},~ {i_{2r+2}} $ are nodes at level $2r+2$ such
that
$\{ {i_{2r+1}}, {i_{2r+2}}\}\in
\mc{N}^*\backslash\mc{M}^*$ and $\{ {i_{-(2r+1)}}, {i_{-(2r+2)}}\}
\in \mc{M}^*\backslash\mc{N}^*$.
Note that, by definition, such paths $P_{\ell}$ for $0\leq \ell \leq
t$ exist since the tree $T_{ i}^t$ has $t+1$ levels and can support
a path of length at most $2t$ as defined above.
Now consider the path $P_t$ of length $2t$. It is an alternating
path on the computation tree with edges from $\mc{M^*}$ and $\mc{N}^*$.
Let us refer to the edges of $\mc{M^*}$ ($\mc{N}^*$) as the
$\mc{M^*}$-edges ($\mc{N}^*$-edges) of $P_t$.

We will now modify the perfect tree-$\mrm{b}$-matching $\mc{N}^*$ by
replacing all $\mc{N}^*$-edges of $P_t$ with their complement in
$P_t$ ($\mc{M^*}$-edges of $P_t$). It is straightforward that this
process produces a new perfect tree-$\mrm{b}$-matching $\mc{N}'$ in
$T_{ i}^t$. 


Let us assume, for the moment, the following lemma:

\begin{lemma}\label{lem:switch}
The weight of the perfect tree-$\mrm{b}$-matching $\mc{N}'$ is strictly less
than that of $\mc{N}^*$ on $T_{ i}^t$.
\end{lemma}
This completes the proof of Lemma \ref{lem:tree=graph} since Lemma
\ref{lem:switch} shows that $\mc{N}^*$ is not the minimum weight
perfect tree-$\mrm{b}$-matching on $T_{ i}^t$, leading to a
contradiction.\enp
\end{proof}
Now, we provide the proof of Lemma \ref{lem:switch}.

\begin{proof}[Proof of Lemma \ref{lem:switch}] It suffices to show that the total
weight of the $\mc{N}^*$-edges of $P_t$ is more than the total
weight of $\mc{M^*}$-edges of $P_t$. For each vertex $ {i_r}\in
P_t$ consider the value $y_{i_r}^*$ from the optimum solution to the dual LP
(\ref{eq:LP-Relax}). Using the inequality $w_{ij}\leq y_i^*+y_j^*$ for
edges of $\mc{M}^*$, we obtain:
\begin{eqnarray}
\sum_{\{i,j\}\in P_t\cap \mc{M}^*}w_{ij}& \leq &
\left(\sum_{r=-t}^{t}y_{i_r}^*\right) -
y_{i_{(-1)^tt}}^*-k_1\ep\label{eq:sum-w-opt-graph}
\end{eqnarray}
where $k_1$ is the number of $\mc{M}^*$-edges of $P_t$ that belong to
$S$, {i.e.,} the number of $\mc{M}^*$-edges of
$P_t$ endowed with the strict
inequality $w_{ij}\leq y_i^*+y_j^*$, with a gap of at least $\ep$. On
the other hand, using the inequality $w_{ij}\geq y_i^*+y_j^*$ for edges
of $\mc{N}^*$ we have:
\begin{eqnarray}
\sum_{\{i,j\}\in P_t\cap \mc{N}^*}w_{ij}& \geq &
\left(\sum_{r=-t}^{t}y_{i_r}^*\right) -
y_{i_{(-1)^{t+1}t}}^*+k_2\epsilon\label{eq:sum-w-opt-tree}
\end{eqnarray}
where now $k_2$ is number of $\mc{N}^*$-edges of $P_t$ that belong to
$S$, or equivalently the number of times the inequality $w_{ij}\geq
y_i^*+y_j^*$ is strict with a gap of at least $\ep$. One finds
\begin{eqnarray}
\sum_{\{i,j\}\in P_t\cap \mc{N}^*}w_{ij}
-\sum_{\{i,j\}\in P_t\cap \mc{M}^*}w_{ij}
& =& y_{i_{(-1)^tt}}^*- y_{i_{(-1)^{t+1}t}}^*+(k_1+k_2)\epsilon
\nonumber\\
&\stackrel{(a)}{\geq}& (k_1+k_2)\epsilon-2L\stackrel{(b)}{\geq} (k_1+k_2)\epsilon-2L\stackrel{(c)}{>}0\label{eq:tree-minus-graph}
\end{eqnarray}
where $(a)$ uses definition of $L$ from Section \ref{sec:Def-Prob-Stat} and $(b)$ uses the fact
that for all $i,j:~\la_{ij}^*\geq 0$. The main step is $(c)$, which
uses Lemma \ref{lem:path-2n} as follows. Path $P_t$ has length $2t$,
and each continuous piece of it with length $2n$ has a projection to
{the} graph $G$ which satisfies
{the}  conditions of Lemma \ref{lem:path-2n}.
This means the path has at least one edge from the set $S$. Thus
$(k_1+k_2)\geq \frac{2t}{2n}>\frac{2L}{\epsilon}$. This completes
the proof of Lemma \ref{lem:switch}.\enp
\end{proof}

\subsection{Independence from Initial Conditions.}\label{sec:indep-initial}
We would like to point out that changing the initial condition for
the messages in step (2) of Sync-BP to any arbitrary values
does not change the convergence and correctness of algorithm Sync-BP.
The only effect of initial condition is
on the number of iterations needed for convergence.
Theorem \ref{thm:main} remains true by \emph{re-defining}
$L$ {according to}: $L=\max_{1\leq i\leq n}~|y_i^*|+\max_{\{i,j\}\in E}~|m_{i\to j}(0)|$.
{This follows} because, by changing {the} initial condition,
the algorithm Sync-BP
runs over a slightly modified computation tree. {The new computation
tree is almost}
the same computation tree as $T_{ i}^t$, except {that} the leaf edges of
the tree have arbitrary weights and not $w_{ij}$'s from $G$.
In the proof of Lemma \ref{lem:tree=graph}, the only place where the
weight of leaf edges appears is the inequality
$(a)$ in equation (\ref{eq:tree-minus-graph}),
which will be satisfied by new definition of $L$.

\subsection{Sync-BP is Correct When $\epsilon$ is Not Well-Defined}\label{sec:epsilon=0}
Recall from the discussion in Section \ref{sec:Def-Prob-Stat} that, if for all edges
$\{i,j\}\in E$ the equality $w_{ij}=y_i^*+y_j^*$ holds, then $\epsilon$ is not well defined.
In this section we show that these rare cases do not cause any trouble. We will show that
the condition $t>\frac{2nL}{\epsilon}$ in the main theorem can be replaced by $t>n$.
This is shown by proving the following lemma instead of Lemma \ref{lem:tree=graph}.
\begin{lemma}\label{lem:tree=graph-epsilon=0}
{If the LP relaxation (\ref{eq:LP-Relax}) has no fractional solution, then},
for any
vertex $ i$ of $G$ and for any $t>n$, the set of
edges that are adjacent to root $ i$ in $\mc{N}^*(T_{ i}^t)$ are
exactly those edges that are connected to $ i$ in $M^*$.
\end{lemma}
\begin{proof}
The proof is similar to the proof of Lemma \ref{lem:tree=graph}. If
 after iteration $t$, {the} claim of the Lemma \ref{lem:tree=graph-epsilon=0}
 does not hold, then the alternating path $P_t$ can be constructed as before.
 Now since {the}
 length of $P_t$ is greater than $2n$, one can use the technical
 Lemma \ref{lem:path-2n} for
the projection of the path $P_t$ onto $G$ to show that the
strict inequality $|w_{ij}-y_i^*-y_j^*|>0$ happens for at least one
edge. This contradicts the above assumption at the beginning of
the Section. Therefore Lemma \ref{lem:tree=graph-epsilon=0} is true.
\enp
\end{proof}


\section{Extension to Possibly Non-Perfect $\mrm{b}$-Matchings}\label{sec:non-perfect}
In this section we show that the algorithm and the results
of the previous sections can be easily generalized to
the case of $\mrm{b}$-matchings (subgraphs $H$ of $G$
such that degree of each vertex $i$ in $H$ is {\emph{at most}} $b_i$).
Let $U(H)\subset V$ be the set of \emph{unsaturated} vertices of $G$
(vertices $i\in V$ such that $deg_H(i)<b_i$).
Similar to Section \ref{sec:Def-Prob-Stat}, the minimum weight
$\mrm{b}$-Matching ($\mrm{b}$-MWM), $H^*$,
is the $\mrm{b}$-Matching such that
\[
H^*=\textrm{argmax}_{H\in \M_G(\mrm{b})}\ W_{H}.
\]
Note that $H^*$ does not include any edge with positive weight
because removing such edges from $H^*$ reduces its weight while
keeping it a $\mrm{b}$-matching. Therefore in this section we
assume that for all $\{i,j\}\in E:~~w_{ij}\leq 0$. The LP
relaxation is slightly different from before:
\begin{equation}
\begin{array}{rcclcrccl}
&&&&&&&&\\
\textrm{min }&&\sum_{\{i,j\}\in E}x_{ij}w_{ij}&&|&\textrm{max }&&\sum_{i=1}^n -b_iy_i-\sum_{{\{i,j\}\in E}} \la_{ij}&\\
\textrm{subject to}&&&&|&\textrm{subject to}&&&\\
&&\sum_{j\in N(i)} x_{ij} \leq b_i&\forall~~ i&|&&&w_{ij}+\la_{ij}\geq -y_i - y_j&\forall~~ \{i,j\}\in E\\
&&0\leq x_{ij} \leq 1&\forall~\{i,j\}\in E&|&&&\la_{ij}\geq0&\forall~\{i,j\}\in E\\
&&&&|&&&&\\
&&&&|&&&&\\
&&\textrm{Primal LP}&&|&&&\textrm{Dual LP}&\\
\end{array}
\label{eq:2nd-LP-Relax}
\end{equation}
\vspace{2mm}
Complementary slackness now reads, for all $\{i,j\}\in E$:  $x_{ij}^*(w_{ij}+\la_{ij}^*+y_i^*+y_j^*)=0$,  $(x_{ij}^*-1)\la_{ij}^*=0$ and
for all $i\in V$: $(\sum_{j\in N(i)}x_{ij}-b_i)y_i^*=0$.

Similarly to Section \ref{sec:Def-Prob-Stat}, we can write the following
modified complementary slackness condition using the
{fact that the LP relaxation has no fractional solution}:
\begin{itemize}
\item[(CS'-i)] For all $\{i,j\}\in H^*;~~~w_{ij}+\la_{ij}^*+y_i^*+y_j^*=0$.
\item[(CS'-ii)] For all $\{i,j\}\notin H^*;~~~\la_{ij}^*=0$.
\item[(CS'-iii)] For all $i\in U(H^*);~~~y_i^*=0$.
\end{itemize}
Let $S'$ be set of those edges in $G$ for
which $|w_{ij}+y_i^*+y_j^*|>0$. We will assume the minimum gap is
$\epsilon'$. That is
\[0<\epsilon' = \min_{\{i,j\}\in S}~|w_{ij} + y_i^*+y_j^*|.\]
The quantity $L'$ is defined {similarly} to $L$ by $L' = \max_{1\leq i\leq n}~|y_i^*|$.

Now we can present the modified algorithm Sync-BP for finding $\mrm{b}$-MWM in $G$:

\vspace{.3in} \hrule \vspace{.1in} \noindent{\bf Algorithm Sync-BP(2).}
\vspace{.1in} \hrule \vspace{.1in}
\begin{itemize}
\item[(1)] At times $t=0,1,\ldots$, each vertex sends
{real-valued} messages
to each of its neighbors. {The} message of $i$ to
$j$ at time $t$ is denoted by $m_{i\to j}(t)$.

\item[(2)] Messages are initialized by $m_{i\to j}(0)
= w_{ij}$ for all $\{i,j\}\in E$.

\item[(3)] For $t \geq 1$, messages in iteration $t$ are obtained from
messages {in} iteration $t-1$ recursively as follows:
\begin{eqnarray}
\forall~\{i,j\}\in E:~~~~m_{i\to j}(t) & = & w_{ij} -
\min\left(0,b_i^{th}\textrm{-min}_{\ell\in N(i)\backslash\{j\}}\bigg[
m_{\ell\to i}(t-1)\bigg]\right) \label{l:recursesim2}
\end{eqnarray}
where $k^{th}$-min$(A)$ {denotes} the $k^{th}$ minimum\footnote{Here $b_i^{th}\textrm{-min}_{\ell\in N(i)\backslash\{j\}}$ is defined to be $0$ if $deg_G(i)=b_i$.} of set A.

\item[(4)] The estimated $\mrm{b}$-MWM at the end of iteration $t$ is $H(t)=\cup_{i=1}^nF_i(t)$ where $F_i(t)=\big\{\{i,j_1\},\ldots,\{i,j_{c_i}\}\big\}$ {is} such that $m_{{j_\ell}\to i}(t) < 0$ for all $1\leq\ell\leq c_i$, i.e., choose
 edges that transfer negative messages to $i$.

\item[(5)] Repeat (3)-(4) {until} $H(t)$ converges.
\end{itemize}
\vspace{.1in} \hrule \vspace{.1in}
The results for $\mrm{b}$-matchings generalize as follows:
\begin{theorem}\label{thm:main-non-perf}
{Assume that the LP relaxation (\ref{eq:2nd-LP-Relax}) has no fractional solution.}
{Then} the algorithm Sync-BP(2) converges to $H^*$ after at most
$\lceil \frac{4nL'}{\epsilon'}\rceil$ iterations.
\end{theorem}

The proof of Theorem \ref{thm:main-non-perf} is similar to the one of
Section \ref{sec:analysis} with the following modifications:

\begin{enumerate}
\item The computation tree $T_i^t$ and $\mrm{b}$-TMWM are defined as before,
while  Lemma \ref{lem:m=n} is slightly modified.
A careful analysis of $W^+$ and $W^-$ for the tree-$\mrm{b}$-matchings
yields equations (\ref{l:recursesim2}) for finding $\mrm{b}$-TMWM in
the computation tree. This is how the new equations are obtained.

\item The technical lemma from Section \ref{subsec:tech-lemma} is still
true and its proof does not change because the definition of alternating
{paths} is preserved and
{because} all cycles involved in the proof turn out to be
adjacent to exactly one edge of $H^*$.

\item {The proof} of Lemmas \ref{lem:tree=graph} and \ref{lem:switch}
should be slightly modified. {In particular, the}
alternating path $P_t$ can be different: One can
show that if {the} $\mrm{b}$-TMWM $\mc{N}^*(T_i^t)$ and the tree-$\mrm{b}$-matching
$\mc{H}^*$ choose different sets of edges at {the} root $i$, then an alternating path
can be constructed as before in $T_i^t$ {which} includes the root $i$. But endpoints
of this alternating path $P_t$ are either leaves of $T_i^t$ or vertices inside
$T_i^t$ {which} have labels from $U(H^*)$ (are un-saturated in $G$ by $H^*$).
{In the case in which there is} at least one leaf as an endpoint of $P_t$, the same
argument as equation (\ref{eq:tree-minus-graph}) in Section
\ref{subsec:comp-tree=graph} can be used since length of $P_t$ is at
least $t$. This shows $(k_1+k_2)\geq \frac{t}{2n}>\frac{2L'}{\epsilon'}$.
But {in the case in which both endpoints of $P_t$ are non-leaf vertices of
 the computation tree, then} using condition (CS'-iii), the analogous version of equation
(\ref{eq:tree-minus-graph}) is as follows:
\begin{eqnarray}
\sum_{\{i,j\}\in P_t\cap \mc{N}^*}w_{ij}
-\sum_{\{i,j\}\in P_t\cap \mc{H}^*}w_{ij}
& =& (k_1+k_2)\epsilon\label{eq:tree-minus-graph-2}.
\end{eqnarray}
Now all that is needed is to show $k_1+k_2>0$. We will show this by
{the} following extension of technical Lemma \ref{lem:path-2n}.


\begin{lemma}\label{lem:any-alt-path}
{Assume that the LP relaxation (\ref{eq:2nd-LP-Relax}) has no fractional solution}.
Then for any alternating path $P$ with endpoints from the set $U(H^*)$, there
{exists} an edge $\{i,j\}\in P$ such that the inequality
$|w_{ij}+y_i^*+y_j^*|>0$ holds. That is, $P\cap S'\neq\emptyset$.
\end{lemma}
\begin{proof}
For paths $P$ with length at least $2n$,
{we can use Lemma \ref{lem:path-2n}},
so there is nothing to do. If a subgraph generated
by $P$ includes at least two cycles, then the same argument as
in {the} proof of Lemma \ref{lem:path-2n} can be used. Therefore
we can assume $P$ intersects itself at most once. So $P$ can
be written as a union $C\cup P_1$ where $C$ is an odd simple
alternating cycle and $P_1$ is a simple alternating path
(either $C$ or $P_1$ can be empty, but not at the same time).
Next, one can define a different solution $\mrm{x}'$ to
{the} LP
(\ref{eq:2nd-LP-Relax}) {which} has the same cost as
$\mrm{x}^*$ by defining $\mrm{x}'=1-\mrm{x}^*$ on path $P_1$
and setting $\mrm{x}'$ equal to $0.5$ on $C$. $\mrm{x}'$ will
still be a feasible solution since the endpoints of $P_1$ are
elements of $U(H^*)$ and the edge adjacent to them in path $P_1$
is not in $H^*$.
This contradicts the {no fractional solution assumption on the LP}.
\enp
\end{proof}
\end{enumerate}


\section{Analysis of the Asynchronous BP}\label{sec:async-BP}
In this section we study the asynchronous version of the BP algorithm.
The update equations are exactly {analogous} to the synchronous
version, but at each time only a subset of the edges are updated
{in} an arbitrary order.
Consider the set $\vec{E}$ of all directed edges in the $G$; i.e.,
 $\vec{E}=\{(i\to j)~~s.t.~~~ i\neq j\in V\}$.
Let $A$ be a sequence $\vec{E}(1),\vec{E}(2),\ldots$ of subsets of the set $\vec{E}$. Then the asynchronous BP algorithm corresponding to the sequence $A$ can be obtained by modifying only the step (3) of the algorithm Sync-BP for the perfect $\mrm{b}$-matchings:
\begin{itemize}
\item[(3)] For $t \geq 1$, messages in iteration $t$ are obtained from messages {in} iteration
$t-1$ recursively as follows:
\[
m_{i\to j}(t) = \left\{
\begin{array}{ll}
 w_{ij} - b_i^{th}\textrm{-min}_{ \ell\in N( i)\backslash\{ j\}}
 \bigg[ m_{ \ell\to  i}(t-1)\bigg]& \textrm{{if}}~ (i\to j)\in\vec{E}(t)\\
 m_{i\to j}(t-1)&\textrm{{otherwise}}
\label{eq:async-update}
\end{array}
\right.
\]
\end{itemize}
\noindent{\bf Note 2.}
This is the most general form of the asynchronous BP and it includes the synchronous version ($\vec{E}(t)=\vec{E}$
for all $t=1,2,\ldots$) {as a special case}. In many applications, a special case
of the asynchronous BP is used for which each set
$\vec{E}(t)$ consists of a single {element.}

We assume that the sequence $A$ of the updates does not have \emph{redundancies}.
That {is},
no edge direction $(i\to j)\in \vec{E}$ is re-updated before
at least one of its {incoming} edge directions ($(\ell\to i)$ for
$\ell\in N(i)\backslash\{j\}$) is updated. More formally, if
$(i\to j)\in \vec{E}(t)\cap\vec{E}(t+s)$ and
$(i\to j)\notin \cup_{r=1}^{s-1}\vec{E}(t+r)$, then at least for
one $\ell\in N(i)\backslash\{j\}$,
we should have $(\ell\to i)\in \cup_{r=1}^{s-1}\vec{E}(t+r)$.

Let us denote the above algorithm by Async-BP. We claim that, if each edge
direction $(i\to j)\in\vec{E}$ is updated $\theta(n)$ times, then the
same result as Theorem \ref{thm:main} can be {proved} here.
That is, let $u(t)$ be the minimum
number of times that an edge direction of the {graph} $G$ appears in the
sequence $\vec{E}(1),\ldots,\vec{E}(t)$; i.e.,
\[u(t) = \min_{(i\to j)\in\vec{E}}\Big(\bigg|\big\{\ell:~~s.t.~~ 1\leq\ell\leq
t~~~\textrm{and}~~~(i\to j)\in\vec{E}(\ell)\big\}\bigg|\Big).\]
From the definition, $u(t)$ is a non-decreasing function
of $t$. {We claim that the
following result holds:}
\begin{theorem}\label{thm:async-b-match}
{Assume that the LP relaxation (\ref{eq:LP-Relax})
has no fractional solution.}
{Then} the algorithm Async-BP
converges to $M^*$ after at most $t$ iterations, provided
$u(t)> \frac{2nL}{\epsilon}$.
\end{theorem}
Before proving the above theorem let us define the notion of
generalized computation tree for the asynchronous version of the BP algorithm.

\subsection{Generalized Computation Tree for the Asynchronous BP}\label{subsec:comp-tree-async}
In order to define the generalized computation tree (GCT) for the asynchronous BP, we
will {begin with} some definitions. For any $(i\to j)\in \vec{E}(t)$,
define $R_{i\to j}^t$ to be the \emph{computation branch} of $i$ to
$j$ at time $t$ {which} is a weighted rooted tree (not necessarily a
balanced rooted tree) and recursively defined according to the
following rules:

(a) {The} root has label $j$.

(b) {The} root has only one child {which} has label $i$.

(c) If $t=0$, then the child $i$ has no child ($R_{i\to j}^0$
is just a single edge $\{i,j\}$).

(d) For $t>0$, if $(i\to j)\notin\vec{E}(t)$ then
$R_{i\to j}^t=R_{i\to j}^{t-1}$. Otherwise the child
$i$ has $deg_G(i)-1$ children {which} have all of labels in the set
$N(i)\backslash\{j\}$
and for any child $r$ of $i$ the subtree that consists of all descendants of $r$ and
the edge $\{r,i\}$ to $i$ is $R_{r\to i}^{t-1}$.
%
%
%

The edge between nodes labeled $i, j$ in the tree is assigned
weight $w_{ij}$ for $1\leq i<j\leq n$.
Now for any vertex $i\in V$ and any $t$,
the GCT $R_i^t$ is a weighted rooted
tree with root $i$ such that all its branches starting
from the root are the computation branches $R_{r\to i}^t$
for all $r\in N(i)$. Since the GCT $R_i^t$
is not necessarily balanced, we will define its \emph{depth}
to be the length of the shortest path from the root $i$
to a leaf and denote it by $d(R_i^t)$.

{Similarly} to the Section \ref{subsec:BP-solv-comp-tree},
we can define the minimum weight perfect
tree-$\mrm{b}$-matching ($\mrm{b}$-TMWPM) for GCT $R_i^t$ and denote it by $\mc{M}^*(R_i^t)$.
Moreover, arguments similar to the ones in the Section \ref{subsec:BP-solv-comp-tree}
show that the algorithm Async-BP is solving the $\mrm{b}$-TMWPM
for GCTs $R_i^t$. In other words, the following
corollary holds:
\begin{corollary}\label{cor:bp-solves-tree-async}
The algorithm Async-BP solves the $\mrm{b}$-TMWPM problem on the GCT. In particular, for each vertex $ i$ of the $G$, the set
$E_i(t)$ that was chosen at the end of iteration $t$ by
Async-BP is exactly the set of $b_i$ edges that are attached to the
root in $\mrm{b}$-TMWPM of $R_{ i}^{t}$.
\end{corollary}

\subsection{Technical Analysis of the Asynchronous BP}
Now we can use the same analysis as in Section
\ref{subsec:comp-tree=graph} to show that
if the depth of the generalized computation tree (GCT) is large enough, then for
any vertex $i$, its neighbors in $M^*$ ($\mrm{b}$-MWPM of $G$) are
exactly those children that are selected in $\mc{N}^*(R_{ i}^{t})$
($\mrm{b}$-TMWPM of $R_{ i}^{t}$). We will show this by relating the
function $u(t)$ to the depth of the GCT
Here is the main lemma {which}
summarizes the above claim:
\begin{lemma}\label{lem:tree=graph-async}
{If the LP relaxation (\ref{eq:LP-Relax}) has no fractional solution, then} for any
vertex $ i$ of  $G$ and for any $t$ such that $u(t)> \frac{2nL}{\epsilon}$,
the set of
edges that are adjacent to root $ i$ in $\mc{N}^*(R_{ i}^t)$ are
exactly those edges that are connected to $ i$ in  $M^*$.
\end{lemma}

The proof of Lemma \ref{lem:tree=graph-async} is similar to
the proof of Lemma \ref{lem:tree=graph}, with the following slight modifications:

(i) One can construct alternating paths $P_\ell$ in the same way
as before for $1\leq \ell\leq d(R_i^t)$.

(ii) The depth of the GCT $R_i^t$ is related to $u(t)$ according to the following lemma:
\begin{lemma}\label{lem:depth=u(t)}
For any vertex $i\in V$ and any $t$, the depth of any computation branch
at time $t$ is at least $u(t)$; i.e., $d(R_{i\to j}^t)\geq u(t).$
\end{lemma}
This tells us $d(R_i^t)\geq u(t)>\frac{2nL}{\epsilon}$.
Applying
this to the path $P_{d(R_i^t)}$, analogously to the use of
Lemma \ref{lem:switch} in the proof of
Lemma \ref{lem:tree=graph}, gives us the proof of
Lemma \ref{lem:tree=graph-async}.
Therefore all that is needed is a proof of Lemma \ref{lem:depth=u(t)}.
\begin{proof}[Proof of Lemma \ref{lem:depth=u(t)}]
The proof {follows easily} by looking at the construction of the computation
branch. Each computation branch $R_{i\to j}^t$ grows at time $t$
if $(i\to j)\in \vec{E}(t)$. {And if this is the case, the depth increases by
at least one due to the ``no redundancy condition'' on the
updating sequence.}  So if each edge is updated at least $u(t)$ times
then the depth of its computation branch grows by at least $u(t)$.
\enp
\end{proof}

Finally we note that the same algorithm as Async-BP and the same result
as Theorem \ref{thm:async-b-match} can be stated and proved for the (possibly non-perfect)
$\mrm{b}$-matchings as well.

\section{Acknowledgements}\label{sec:ack}
We would like to thank L\'{a}szl\'{o} Lov\'{a}sz, Andrea Montanari,
Elchannan Mossel and Amin Saberi for useful discussions.
{This work was
done while Riccardo Zecchina was a Visiting Researcher in the Theory
Group at Microsoft Research, and was supported by the Microsoft Technical Computing
Initiative.}

\vspace{5mm}
\appendix
{\Large \bf Appendix}
\vspace{1mm}

\section{Counterexample}\label{sec:counter-example}
To see that strictness of inequality $w_{ij}\geq y_i^*+y_j^*$ from Section \ref{sec:Def-Prob-Stat} does not hold in general, consider the case of $\mrm{1}$-matching on the
complete graph with four vertices and weights
$w_{12}=w_{24}=w_{34}=a~,~w_{13}=w_{14}=w_{23}=b$ where $b\gg a$. It
is clear that $M^*=\{(1,2),(3,4)\}$ is the unique minimum weight matching.
Consider the following, feasible solutions
to the LP and its dual:
$x_{12}'=x_{34}'=1~,~x_{13}'=x_{14}'=x_{23}'=x_{24}'=0$,  $\mrm{\la}'=\mrm{0}$, and
$\mrm{y}'=(a/2,a/2,a/2,a/2)$.  These solution satisfy the complementary slackness
conditions. Therefore $\mrm{x}^*=\mrm{x}'$ and
$\mrm{\la}^*=\mrm{\la}',~\mrm{y}^*=\mrm{y}'$.}  But for the edge
$(2,4)\notin M^*$ the equality $w_{24}=y_2^*{+}y_4^*$ holds. The same
example can be used for the case of (non-perfect) matching as well.






\end{document}